\documentclass[aps,prb,showpacs,twocolumn,amsmath,amssymb,superscriptaddress]{revtex4}
\usepackage{graphicx}
\usepackage{Jonasmacros}
\usepackage{bm}
\begin{document}
\date{\today}
\title{Imaging spin-inelastic Friedel oscillations emerging from magnetic impurities}
\author{J. Fransson}
\email{Jonas.Fransson@fysik.uu.se}
\affiliation{Department of Physics and Materials Science, Uppsala University, Box 530, SE-751 21\ \ Uppsala}

\author{A. V. Balatsky}
\email{avb@lanl.gov}
\affiliation{Theoretical Division, Los Alamos National Laboratory, Los Alamos, New Mexico 87545, USA}
\affiliation{Center for Integrated Nanotechnology, Los Alamos National Laboratory, Los Alamos, New Mexico 87545, USA}

\begin{abstract}
We consider inelastic scattering of localized magnetic moments coupled with the electrons on the surface. We argue that spin-inelastic transitions of the magnetic impurities generate oscillations at a momentum $k$, corresponding to the inelastic mode, in the second derivative of the current with respect to voltage $d^2I/dV^2$. These oscillations are similar in nature to Friedel oscillations. Inelastic Friedel oscillations, which were previously proposed for spin-unpolarized set-up, is here extended for spin-polarized systems. We propose to use scanning tunneling microscope to measure spin-inelastic scattering generated at the impurity spin by imaging the $d^2I/dV^2$ oscillations on the metal surface.
\end{abstract}
\maketitle

Measurements of inelastic transitions open up a route to investigate the excitation spectrum of physical systems. There has been a growing activity in elucidating inelastic scattering processes in quantum systems using various experimental techniques. An incomplete list includes inelastic neutron \cite{prassides1991,christianson2008} and X-ray \cite{lee2005,saiz2008} scattering, transport through break junctions \cite{park2000,wang2004}, and scanning tunneling microscopy (STM) with spin-polarized (SP-STM) \cite{meier2008,zhou2010} or non spin-polarized tip \cite{stipe1998,hirjibehedin2006,hirjibehedin2007,otte2008,otte2009,balashov2009,khajetoorians2011}.

Surface imaging of scattering states can be performed e.g. by using STM to probe the spatial spectral density variations at a given energy. It is well-known that Friedel oscillations emerge around defects adsorbed onto a surface caused by elastic scattering processes \cite{hasegawa1993,sprunger1997}. Less known is the prediction made by  us earlier that  points to the existence of {\em inelastic Friedel oscillations} emerging from vibrating impurities \cite{fransson2007}. These oscillations were recently demonstrated experimentally for dimers of meta-dichlorobenzene \cite{gawronski2010}. The mechanism for inelastic Friedel oscillations is essentially the same as for conventional oscillations and comes from interference of incoming and outgoing waves that have an energy mismatch given by energy transferred to/from local vibrational mode.

In this paper, we propose a spin-polarized extension of inelastic Friedel oscillations that arise from spin-inelastic transitions. We also propose to use STM to image these oscillations. A local magnetic moment interacting with surface electrons, generates a local spin-polarization in the surface. The spin-inelastic transitions provide another modification to the electronic and magnetic structures of the surface states. This local modification experimentally can be stimulated by adding or removing an energy quantum that corresponds to the inelastic transition energy. Such an energy change causes a change in the electronic structure of the type $\delta N(\bfr,\omega)=\sum_{\alpha\beta}\calQ_{\alpha\beta}(\bfr,\omega)\theta(\omega-\Delta_{\beta\alpha})$, for low temperatures. Here, $\Delta_{\beta\alpha}$ is the transition energy, whereas $\calQ_{\alpha\beta}(\bfr,\omega)$ is a spatial distribution function which depends on the involved states $\ket{\alpha}$, $\ket{\beta}$.

In order to enhance the signature of spin-inelastic scattering effects, and image  spin-inelastic Friedel oscillations, we propose to use quantum corrals constructed out of magnetic atoms, or molecules. Placing STM tip at the center of  quantum corrals will allow to amplify the signal. As the magnetic atoms are coupled through exchange interactions, the magnetic structure of the quantum corral can be engineered to meet specific requirements. When coupled anti-ferromagnetically, one can study qualitative differences in quantum corrals comprising an even or odd number of atoms. Ferromagnetic coupling between the atoms, on the other hand, gives rise to a large collective spin moment that could have its own signatures. In either case of ferro- or antiferromagnetically coupled corrals using STM would allow to image specific spatial fingerprints throughout the interior of the corral.


In order to give an example of the effect we are proposing, consider a collection of general (quantum) spins $\bfS_n=\bfS(\bfr_n)$ located at the positions $\bfr_n$ on a metallic substrate surface. The surface electron density can be modeled by $\Hamil_\text{surf}=\sum_\bfk\dote{\bfk}\cdagger{\bfk}\cc{\bfk}$, whereas the Kondo interaction between the surface electron density and the local spin is given as $\Hamil_K=v_uJ_K\sum_n\bfs(\bfr_n)\cdot\bfS_n$, where $v_u$ is the unit area and $J_K$ is the Kondo exchange parameter, whereas $\bfs(\bfr_n)=\csdagger{\sigma}(\bfr_n)\bfsigma_{\sigma\sigma'}\cs{\sigma'}(\bfr_n)$ with spin indices $\sigma,\sigma'=\up,\down$, $\cs{\sigma}(\bfr)=\int\cs{\bfk\sigma}e^{i\bfk\cdot\bfr}d\bfk/(2\pi)^2$, and the vector $\bfsigma$ of Pauli matrices. It should be noticed that the Kondo interaction only provides the isotropic interaction between the local spin moment and the substrate electrons, whereas the anisotropy is being treated separately, see the discussion below.

We make contact with current STM measurements on local magnetic moments by formulating the tunneling (differential) conductance in terms of the Tersoff and Hamann approach \cite{tersoff1983}, and its generalizations \cite{wortmann2001,fransson2010}.
The tunneling conductance at low temperatures is in this approach given by
\begin{align}
\frac{dI(\bfr,V)}{dV}\sim&
			n(\dote{F}-eV)N(\bfr,\dote{F})
\end{align}
where $\dote{F}$ is the Fermi level of the system in equilibrium. This expression, thus, relates the tunneling conductance to the electronic ($N$) 
structure of the substrate surface, and correspondingly ($n$) for the tip. It is, therefore, sufficient to study the local variations in the spin-polarized surface density of electronic states.

In our calculated examples below, we assume that the localized spin moments can be described in terms of the Hamiltonian
\begin{align}
\Hamil_S=&
	\sum_n\{
		D(S_n^z)^2
		+E[(S^+_n)^2+(S_n^-)^2]/2
		\},
\label{eq-HS}
\end{align}
where the anisotropy fields $D$ and $E$ account for the effective interaction between the localized spin moments and the surface electrons. Notice that this Hamiltonian effectively describes the interactions between the localized spin moment and the surface electrons which give rise the anisotropy of the localized spin. The isotropic interactions are accounted for by the Kondo interation $\Hamil_K$. This model defines $2S+1$ eigenstates $\ket{\alpha}$ and eigenenergies $E_\alpha$, and we introduce the operators $\ddagger{\alpha}$ ($\dc{\alpha}$) which create (destroy) a particle in the state $\ket{\alpha}$. For later use, we also define the spin operators $\tau^i_{\alpha\beta}=\ket{\alpha}\bra{\alpha}S^i\ket{\beta}\bra{\beta}$, $i=x,y,z$.

We employ the model given in Eq. (\ref{eq-HS}) since it has been successfully used to describe single (and multiple) impurities located on metallic surface, see e.g. Refs. \onlinecite{hirjibehedin2006,hirjibehedin2007,otte2008,balashov2009,otte2009,fransson2009}. The anisotropy fields $D$ and $E$ are related to the properties of the interactions between the local adsorbant and the substrate material, and can be fitted to the experiment \cite{hirjibehedin2006,hirjibehedin2007} but also determined through first principles calculations \cite{balashov2009}

In absence of impurities, the substrate surface is assumed to be non-magnetic, however, in presence of the local spins the surface LDOS may become spin-polarized locally around the spins. We account for the scattering off the magnetic impurities by calculating the real space Green function (GF) for the surface electrons using $\bfG(\bfr,\bfr';i\omega)=\int\bfG(\bfk,\bfk';i\omega)e^{i\bfk\cdot\bfr-i\bfk'\cdot\bfr'}d\bfk d\bfk'/(2\pi)^4$, where $\bfG(\bfk,\bfk';i\omega)=\{G_{\sigma\sigma'}(\bfk,\bfk';i\omega)\}_{\sigma\sigma'}$ is the $2\times2$ matrix of the spinor $\Psi(\bfk)=(\cs{\bfk\up}\ \cs{\bfk\down})^T$.

First we construct a {\it bare} GF $\bfG^{(0)}$ which contains the spin-polarization induced by the localized magnetic moments, using the model $\Hamil_\text{surf}+\Hamil_K$. In spirit of scattering theory \cite{fiete2001,franssonNL2010}, we obtain a T-matrix (spin space) formulation
\begin{subequations}
\label{eq-bareGF}
\begin{align}
\bfG^{(0)}(\bfk,\bfk')=&
	\delta(\bfk-\bfk')\bfg(\bfk)
\nonumber\\&
	+\sum_{nm}\bfg(\bfk)e^{-i\bfk\cdot\bfr_n}\bfT(\bfr_n,\bfr_m)\bfg(\bfk')e^{i\bfk'\cdot\bfr_m},
\\
\bfT(\bfr_n,\bfr_m)=&
	\bft(\bfr_n,\bfr_m)\bfV_m,
\end{align}
\end{subequations}
where $\bft^{-1}(\bfr_n,\bfr_m)=\delta(\bfr_n-\bfr_m)-\bfV_n\bfg(\bfr_n-\bfr_m)$, and $\bfg(\bfr-\bfr')=\int\bfg(\bfk)e^{i\bfk\cdot(\bfr-\bfr')}d\bfk/(2\pi)^2$, $\bfg(\bfk;i\omega)=(i\omega-\dote{\bfk})^{-1}$. The scattering potential $\bfV=V_0+\bfsigma\cdot\bfDelta_n(i\omega)$ comprises the spin-independent and spin-dependent contributions $V_0$ and $\bfDelta_n(i\omega)=v_uJ_K\av{\bfS_n}(i\omega)$. 

The spin-inelastic scattering off the localized spin moments that influences the surface electron GF, is accounted for in second order perturbation theory with respect to $v_uJ_K$. We define the self-energy \cite{balatsky2003}
\begin{align}
\Sigma_{\sigma\sigma'}(\bfr_n,\bfr_m;i\omega)=&
	-\frac{(v_uJ_K)^2}{\beta}
	\sum_{\nu ss'}\bfsigma_{\sigma s}\cdot\chi_{nm}(i\nu)\cdot\bfsigma_{s'\sigma'}
\nonumber\\&\times
	G_{ss'}^{(0)}(\bfr_n,\bfr_m;i\omega-i\nu)
\label{eq-exp}
\end{align}
($\beta^{-1}=k_BT$) where the spin-spin GF $\chi_{nm}(z)=\int\eqgr{\bfS_n(t)}{\bfS_m(t')}e^{iz(t-t')}dt'$, and obtain the real space GF
\begin{align}
\bfG(\bfr,\bfr')\approx&
	\bfG^{(0)}(\bfr,\bfr')
\nonumber\\&
			+\sum_{nm}
				\bfG^{(0)}(\bfr,\bfr_n)\bfSigma(\bfr_n,\bfr_m)\bfG^{(0)}(\bfr_m,\bfr').
\label{eq-dressedGF}
\end{align}

Next, we discuss the effects of the spin-inelastic scattering on the surface electrons. For the sake of argument we use the unperturbed surface electron GF, i.e. replace $\bfG^{(0)}$ by $\bfg$ in Eq. (\ref{eq-dressedGF}), to allow for analytical calculations. For non-interacting spin impurities, the spin-spin GF can be written
\begin{align}
\chi_{nm}(i\omega)=&
	\frac{\delta_{nm}}{\beta}
		\sum_{\alpha\beta,\nu}
		\bftau_{\alpha\beta}\bftau_{\beta\alpha}
			G_{n\beta}(i\omega+i\nu)G_{n\alpha}(i\nu)
\end{align}
where $G_{n\alpha}(t,t')=\eqgr{\dc{n\alpha}(t)}{\ddagger{n\alpha}(t')}$. Here, we have written the spin in terms of the eigenstates of Eq. (\ref{eq-HS}) such that the spin GF $G_{n\alpha}(i\omega)=(i\omega-E_{n\alpha})^{-1}$. We find that the retarded form of the self-energy in Eq. (\ref{eq-dressedGF}) can be written (using quadratic dispersion $\dote{\bfk}=\hbar^2k^2/2m$)
\begin{align}
\bfSigma^r(\omega)
\approx&
	\frac{\gamma^2}{2\pi N_0}\sum_{\alpha\beta}\bfsigma\cdot\bftau_{\alpha\beta}\bftau_{\beta\alpha}\cdot\bfsigma
	\biggl\{
		i\pi f(-E_\alpha)f(E_\beta)
\nonumber\\&
		+[f(E_\beta)
		-f(E_\alpha)]
		\biggl[
			\ln\frac{|\omega-E_\alpha+E_\beta|}{W}
\nonumber\\&
			+i\pi f(\omega-E_\alpha+E_\beta)
		\biggr]
	\biggr\},
\label{eq-Sigmaindep}
\end{align}
where $2W\sim1$ eV is the band-width, whereas $\gamma=v_uJ_KN_0$, with the bare surface DOS $N_0=m/\hbar^2$, and $f(x)$ is the Fermi function. Noting that $\bfg^r(\bfr)\approx-iN_0J_0(k|\bfr|)/2$, where $J_0(x)$ is the zeroth order Bessel function of the first kind, whereas $k\equiv|\bfk|=\sqrt{2N_0\omega}$, we find that the substrate LDOS $N(\bfr,\omega)=N_0(\bfr,\omega)+\delta N(\bfr,\omega)$, defined by $N(\bfr,\omega)=-\tr\im\bfG(r,r)/\pi$, where $N_0(\bfr,\omega)=N_0$ whereas
\begin{align}
\frac{\delta N(\bfr,\omega)}{N_0}=&
	\frac{\gamma^2}{\pi}
	J_0^2(k|\bfr-\bfr_0|)
	\sum_{\alpha\beta}
		\bftau_{\alpha\beta}\cdot\bftau_{\beta\alpha}
		\Bigl\{
			f(-E_\alpha)f(E_\beta)
\nonumber\\&
			+[f(E_\beta)-f(E_\alpha)]
			f(\omega-E_\alpha+E_\beta)
		\Bigr\}.
\label{eq-dN}
\end{align}
It is clear form this expression that the amplitude of the inelastic signal scales with the square of the Kondo coupling $J_K$.

\begin{figure}[t]
\begin{center}
\includegraphics[width=.99\columnwidth]{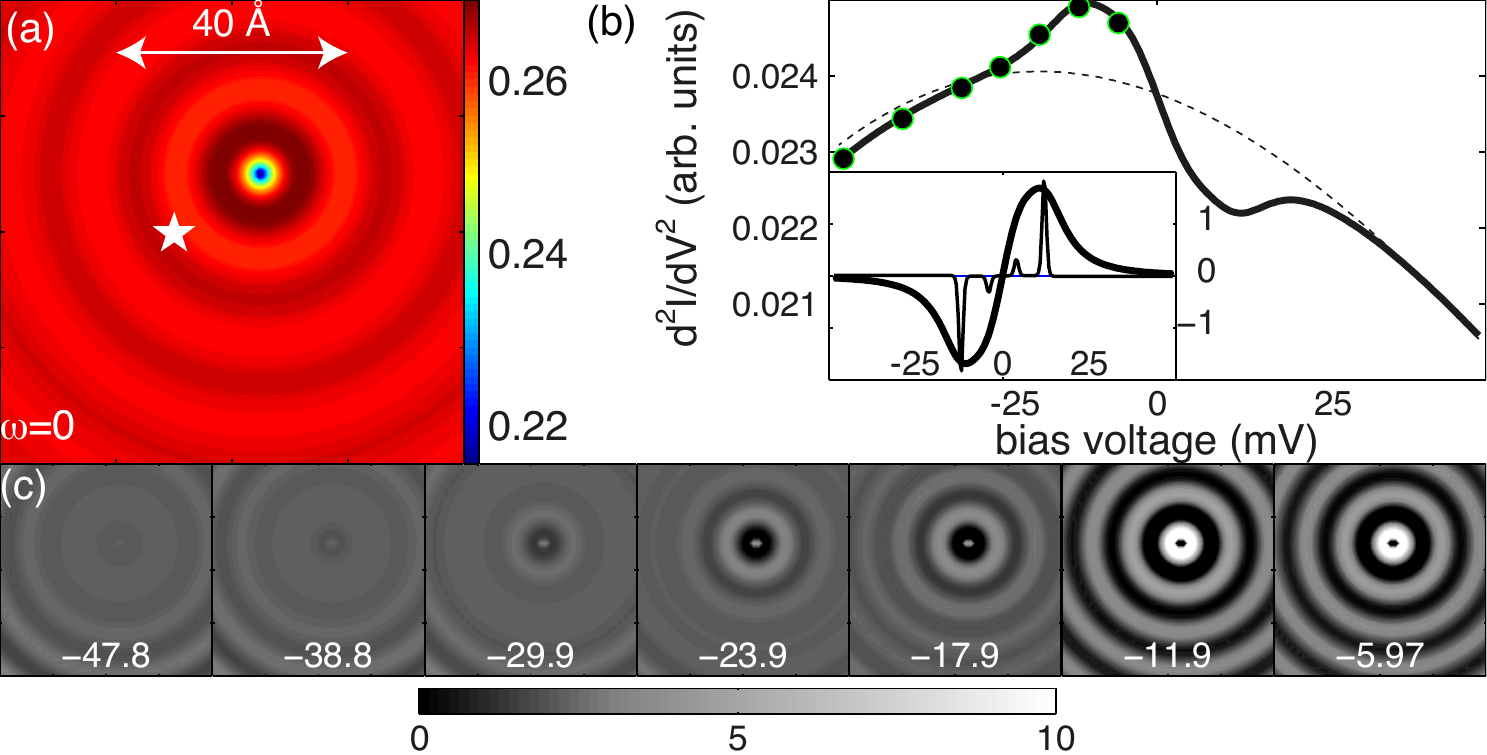}
\end{center}
\caption{(a) DOS ($dI/dV$) map of a single spin $S=1$ impurity adsorbed onto a metallic surface around which concentric Friedel oscillations emerge, calculated using the GF defined in Eqs. (\ref{eq-bareGF}) and (\ref{eq-dressedGF}). (b) IETS ($d^2I/dV^2$) spectrum, with (solid) and without (dashed) spin-inelastic scattering, recorded at a distance $1.8$ nm from the impurity (star in panel (a)). The inset shows the IETS spectrum at the defect, where the two plots correspond to the broadened (bold) and non-broadened (faint) spin states. (c), IETS maps of the system different energies, marked by bullets in panel (b). Here, $D=-10$ meV, $E\sim|D|/5$, $T\sim4$ K.}
\label{fig-N1}
\end{figure}

Feeding the energy $\omega=E_\alpha-E_\beta$ into the system by means of e.g. the bias voltage, stimulates the inelastic spin transition $\ket{\alpha}\bra{\beta}$, and the onset of the inelastic scattering generates an abrupt change in the surface LDOS. The expression in Eq. (\ref{eq-dN}), moreover, shows that the onset of the spin-inelastic scattering generates spatial variations in the charge density emerging from the localized magnetic moment, referred to as spin-inelastic Friedel oscillations, in analogy to previously introduced inelastic Freidel oscillations emerging from vibrational defects \cite{fransson2007,gawronski2010}. The charge density variations are modulated by the momentum $k$. Generally, inelastic scattering is not a Fermi surface effect but rather pinned to the momentum $k$, for which reason one should expect a varying wavelength of the Friedel oscillations emerging from the impurity depending on the energy of the specific inelastic transition. The LDOS connects with current inelastic electron tunneling spectroscopy (IETS) measurements by noting that the signal $d^2I(\bfr,V)/dV^2\propto\partial N(\bfr,\omega)/\partial\omega$. For low temperatures, the derivative $-df(\omega-E_\alpha+E_\beta)/d\omega\rightarrow\delta(\omega-E_\alpha+E_\beta)$, which indicates the possibility to image inelastic Friedel oscillations in a narrow range of energies around the inelastic transfer energy $E_\alpha-E_\beta$.

In Fig. \ref{fig-N1} (a), we plot the local DOS of the surface electrons interacting with a localized $S=1$ spin moment adsorbed onto the surface, pertinent for e.g. Co/Pt(111) \cite{balashov2009}, around which elastic Friedel oscillations emerge in the surface DOS. For the calculations we have used the full electronic GFs prescription as defined in Eqs. (\ref{eq-bareGF}) and (\ref{eq-dressedGF}). Positioning the STM tip at the point marked by a star in panel (a), we plot in Fig. \ref{fig-N1} (b) the IETS ($\partial_\omega N(\bfr_{\tip},\omega)$) spectrum for the perturbed (solid) and unperturbed (dashed) surface. Here, we have added a phenomenological Lorentzian broadening ($\sim7.5$ meV) of the spin states in order to capture the behavior of the IETS spectrum observed in Ref. \cite{balashov2009}. The broadening has been estimated the by fitting the shape of the IETS spectrum at the defect position to experiments \cite{fransson2009}, see inset of Fig. \ref{fig-N1} (b) (bold). The faint plot in the inset of Fig. \ref{fig-N1} shows the IETS spectrum for the spin in the atomic limit.

The setup, thus, demonstrates the possibility to remotely record the inelastic signatures emerging from the scattering center, due to its propagation over the surface via the spin-inelastic Friedel oscillations.

The spatial characteristics is expected to vary significantly with the energy, which indeed can be seen in Fig. \ref{fig-N1} (c), where we plot IETS maps for a few energies corresponding to the energies in Fig. \ref{fig-N1} (b). As is indicated in the IETS spectrum, no essential spatial structure is found for energies far off the inelastic transition energies. In fact, since the IETS spectrum is vanishingly small for energies off the inelastic transition energies, the spatial IETS maps are expected to be nearly equal to the corresponding bare maps. By a comparison between the IETS maps for the different energies, it is clear that the localized moment generates a spatial response, i.e. spin-inelastic Friedel oscillations, for probe energies close to the inelastic transition energies.

The magnetic structure $\bfM(\bfr,\omega)$ emerging from the local spin moment, is in this simplified example reduced to a simple spatially varying spin-polarization $\bfM(\bfr,\omega)=M_z(\bfr,\omega)\hat{\bf z}$ of the surface electrons, which can be calculated from $M_z(\bfr,\omega)=\sum_\sigma\sigma_{\sigma\sigma}^zN_\sigma(\bfr,\omega)$. For weak coupling between the localized spin and the electron medium assumed here, the spin-polarization is negligible.

\begin{figure}[t]
\begin{center}
\includegraphics[width=.99\columnwidth]{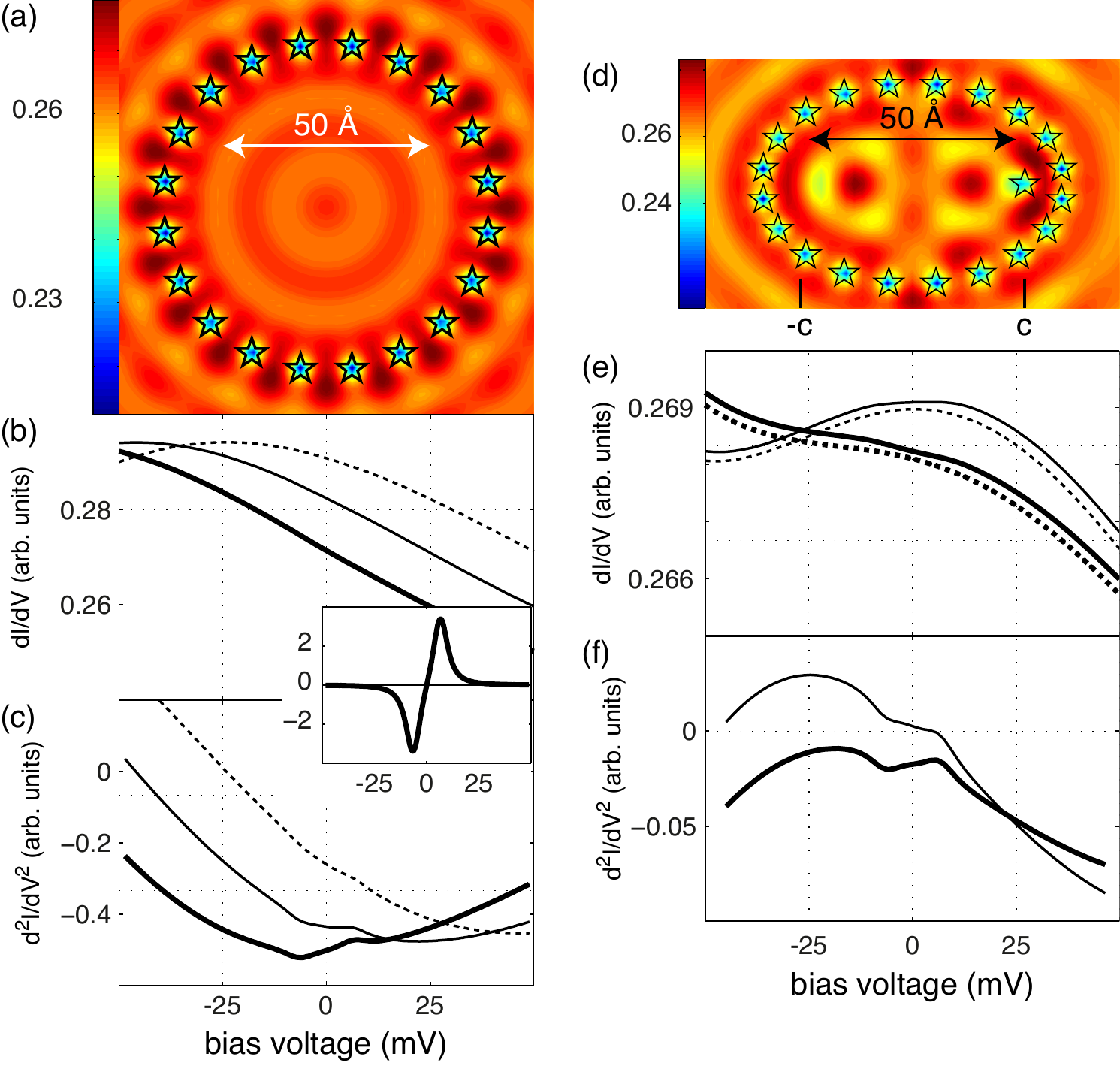}
\end{center}
\caption{(a), (d) $dI/dV$ map of a 20 atom circular and elliptical quantum corral (atomic positions marked by pentagons), respectively, for independent $S=3/2$ spins anti-ferromagnetically coupled, (b) $dI/dV$ calculated at the center of circular corrals for radii $R=39.75+\{0.2,1.2,2.2\}$ \AA, and (c) corresponding IETS ($d^2I/dV^2$) spectrum. Inset shows the IETS for a single $S=3/2$ defect. (e) and (f) $dI/dV$ and $d^2I/dV^2$, respectively, of the corral in (d) calculated at $(x,y)=(-c,0)$ with (bold) and without (faint) an adatom at $(x,y)=(c,0)$. Ellipse in (d) given by $R^2=(x/a)^2+(y/b)^2$, with $R=24.25$ \AA, and $a/b=1.5$. Here, $D=-3.25$ meV, $E=0$, and $T=4$ K.}
\label{fig-N20}
\end{figure}

We now consider quantum corrals comprised of magnetic atoms and implement the theoretical framework introduced above. In particular, we focus on circular geometry and consider the resulting electronic structure for independent spin moments. In Fig. \ref{fig-N20} (a) and (d) we plot  the Fermi level spectral density of circular and elliptical quantum corrals comprising 20 independent $S=3/2$ atoms anti-ferromagnetically coupled. The structure in the spectral density is caused by the confinement imposed by the corral. The corresponding $dI/dV$ calculated at the center of circular corrals with different radii, using a level broadening of  $\sim 5$ meV \cite{balashov2009,fransson2009}, is shown in Fig. \ref{fig-N20} (b). Due to the voltage dependent corral $dI/dV$, we expect $d^2I(\bfr_c,V)/dV^2$ to be non-zero also for voltages far off the spin-inelastic modes, which is shown in Fig. \ref{fig-N20} (c).

For the elliptic corral, in Fig. \ref{fig-N20} (e) we show the $dI/dV$ at the focal point $-c$, c.f. labels in Fig. \ref{fig-N20} (d), for cases with (bold) and without (faint) a spin defect at the focus $c$. It is clear that the electronic structure changes slightly due to the additional defect. More important is that the $dI/dV$ becomes distorted near zero bias voltage due to the spin-inelastic scattering, and those distortions are more clearly seen in Fig. \ref{fig-N20} (f), where the corresponding IETS spectra is plotted. Despite the energy variations of the surface electron density, which accordingly are also picked up in the IETS signal, the spin-inelastic contribution provides a significant distortion of the signal. Our calculations performed for the quantum corrals, hence, provide clear demonstrations that the inelastic scattering should be remotely detectable also within systems with more complicated electronic structures.

The slow energy dependence of the underlying DOS implies that its energy derivative is small, which leads to that the spatial signatures in the IETS maps are correspondingly small for energies sufficiently far off the inelastic transition energy. Close to the inelastic transition energies, on the other hand, we expect to be able to detect the spatial variations in the spectral density, analogous to the maps shown in Fig. \ref{fig-N1} (c).

We have demonstrated theoretically, that it should be possible to image the response to spin-inelastic transitions using STM for IETS measurements. Scattering off the local spin (or magnetic) moment modifies the DOS locally around the impurity, and for specific energies corresponding to the inelastic transition energies, additional modification of the local DOS is expected. The inelastic signatures can be identified as sharp peak/dip features near the inelastic transition energies. Performing IETS measurements will reveal a spatially modulated spectral density near the inelastic modes. While measurements on single magnetic impurities should be sufficient in order to resolve the inelastic Friedel oscillations, we suggest that the corresponding signatures can be substantially enhanced by engineered quantum structures, e.g. quantum corrals.

JF acknowledges support from the Swedish Research Council (622-2007-562). Work at Los Alamos was supported by US DOE, BES and LDRD funds.


\begin{thebibliography}{20}
\bibitem{prassides1991} K. Prassides, J. Tomkinson, C. Christides, M. J. Rosseinsky, D. W. Murphy, R. C. Haddon, Nature {\bf 354}, 462-463 (1991).
\bibitem{christianson2008} A. D. Christianson, E. A. Goremychkin, R. Osborn, S. Rosenkranz, M. D. Lumsden, C. D. Malliakas, I. S. Todorov, H. Claus, D. Y. Chung, M. G. Kanatzidis, R. I. Bewley, and T. Guidi, Nature, {\bf 456}, 930 (2008).

\bibitem{lee2005} S. K. Lee, P. J. Eng, H. -K. Mao, Y. Meng, M. Newville, M. Y. Hu, J. Shu, Nat. Mat., {\bf 4}, 851 (2005).
\bibitem{saiz2008} E. Garc'a Saiz, G. Gregori, D. O. Gericke, J. Vorberger, B. Barbrel, R. J. Clarke, R. R. Freeman, S. H. Glenzer, F. Y. Khattak, M. Koenig, O. L. Landen, D. Neely, P. Neumayer, M. M. Notley, A. Pelka, D. Price, M. Roth, M. Schollmeier, C. Spindloe, R. L. Weber, L.  van Woerkom, K. W\"unsch, and D. Riley, Nat. Phys., {\bf 4}, 940 (2008).

\bibitem{park2000} H. Park, J. Park, A. K. L. Lim, E. H. Anderson, A. P. Alivisatos, and P. L. McEuen, Nature, {\bf 407}, 57 (2000).
\bibitem{wang2004} W. Wang, T. Lee, I. Kretzschmar, and M. A. Reed, Nano Lett. {\bf 4}, 643 (2004).

\bibitem{meier2008} F. Meier, L. Zhou, J. Wiebe, and R. Wiesendanger, Science, {\bf 320}, 82 (2008).
\bibitem{zhou2010} L. Zhou, J. Wiebe, S. Lounis, E. Vedmedenko, F. Meier, S. Bl\"ugel, P. H. Dederichs, and R. Wiesendanger, Nat. Phys. {\bf 6}, 187 (2010).

\bibitem{stipe1998} B. C. Stipe, M. A. Rezaei, and W. Ho, Science, {\bf 280}, 1732 (1998).
\bibitem{hirjibehedin2006} C. F. Hirjibehedin, C. P. Lutz, and A. J. Heinrich, Science, {\bf 312}, 1021 (2006).
\bibitem{hirjibehedin2007} C. F. Hirjibehedin, C. -Y. Lin, A. F. Otte, M. Ternes, C. P. Lutz, B. A. Jones, and A. J. Heinrich, Science, {\bf 317}, 1199 (2007).
\bibitem{otte2008} A. F. Otte, M. Ternes, K. von Bergmann, S. Loth, H. Brune, C. P. Lutz, C. F. Hirjibehedin, and A. J. Heinrich, Nat. Phys. {\bf 4}, 847 (2008).
\bibitem{otte2009} A. F. Otte, M. Ternes, S. Loth, C. P. Lutz, C. F. Hirjibehedin, and A. J. Heinrich, Phys. Rev. Lett. {\bf 103}, 107203 (2009).
\bibitem{balashov2009} T. Balashov, T. Schuh, A. F. Tak\'acs, A. Ernst, S. Ostanin, J. Henk, I. Mertig, P. Bruno, T. Miyamachi, S. Suga, and W. Wulfhekel, Phys. Rev. Lett. {\bf 102}, 257203 (2009).
\bibitem{khajetoorians2011} A. A. Khajetoorians, S. Lounis, B. Chilian, A. T. Costa, L. Zhou, D. L. Mills, J. Wiebe, and R. Wiesendanger, Phys. Rev. Lett. {\bf 106}, 037205 (2011).

\bibitem{hasegawa1993} Y. Hasegawa and Ph. Avouris, Phys. Rev. Lett. {\bf 71}, 1071 (1993).
\bibitem{sprunger1997} P. T. Sprunger, L. Petersen, E. W. Plummer, E. L\ae gsgaard, and F. Besenbacher, Science, {\bf 275}, 1764 (1997).

\bibitem{fransson2007} J. Fransson and A. V. Balatsky, Phys. Rev. B, {\bf 75}, 195337 (2007).
\bibitem{gawronski2010} H. Gawronski, J. Fransson, and K. Morgenstern, Nano Lett. {\bf 11}, 2720 (2011).

\bibitem{tersoff1983} J. Tersoff and D. R. Hamann, Phys. Rev. Lett. {\bf 50}, 1998 (1983).
\bibitem{wortmann2001} D. Wortmann, S. Heinze, Ph. Kurz, G. Bihlmayer, and S. Bl\"ugel, Phys. Rev. Lett. {\bf 86}, 4132 (2001).

\bibitem{fransson2010} J. Fransson, O. Eriksson, and A. V. Balatsky, Phys. Rev. B, {\bf 81}, 115454 (2010).
\bibitem{fransson2009} J. Fransson, Nano Lett. {\bf 9}, 2414 (2009).

\bibitem{fiete2001} G. A. Fiete, J. S. Hersch, E. J. Heller, H. C. Manoharan, C. P. Lutz, and D. M. Eigler, Phys. Rev. Lett. {\bf 86}, 2392 (2001).
\bibitem{franssonNL2010} J. Fransson, H. C. Manoharan, and A. V. Balatsky, Nano Lett. {\bf 10}, 1600 (2010).

\bibitem{balatsky2003} A. V. Balatsky, Ar. Abanov, and J. -X. Zhu, Phys. Rev. B, {\bf 68}, 214506 (2003).

\end{thebibliography}
\end{document}